\documentclass[12pt]{article}

\usepackage{graphicx}
\usepackage{amsmath}
\usepackage{amssymb}
\usepackage{amsthm}
\usepackage{latexsym}
\usepackage{mathrsfs}
\usepackage{enumerate}
\usepackage{ifthen}

\theoremstyle{definition}

\newtheorem*{result}{Theorem}

\newcommand{\stateset}{\mbox{$\mathscr{S}$}}

\newcommand{\eidoset}{\mbox{$\mathscr{E}$}}

\newcommand{\infoset}{\mbox{$\mathscr{I}$}}
\newcommand{\mechset}{\mbox{$\mathscr{M}$}}

\newcommand{\uniformset}{\mbox{$\mathcal{U}$}}

\newcommand{\fproc}[2]{\left \langle #1,#2 \right \rangle}

\newcommand{\entropy}{\mathbb{S}}
\newcommand{\bitstate}{I_{\mathrm{b}}}

\newcommand{\avg}[1]{\left \langle #1 \right \rangle}
\newcommand{\boxstate}[2]{b^{#1}_{#2}}

\newcommand{\leftorright}{\rightleftharpoons}

\title{Entropic probability and context states}
\date{\today}

\author{Benjamin Schumacher\thanks{Corresponding author:  Department of Physics,
	Kenyon College, Gambier, OH 43022 USA. E-mail schumacherb@kenyon.edu} 
	\\ Department of Physics, Kenyon College\\
	\and Michael D. Westmoreland\\Department of Mathematics, Denison University}

\begin{document}

\maketitle

\begin{abstract}
	In a previous paper, we introduced an axiomatic system for information
	thermodynamics, deriving an entropy function that includes both thermodynamic
	and information components.  From this function we derived an entropic
	probability distribution for certain uniform collections of states.  Here we
	extend the concept of entropic probability to more general collections, 
	augmenting the states by reservoir and context states.  This leads to an
	abstract concept of free energy and establishes a relation between
	free energy, information erasure, and generalized work.
\end{abstract}

\section{Introduction}

In \cite{AxiomaticInfoThermo}, we developed an axiomatic system
for thermodynamics that incorporated information as a fundamental concept.
This system was inspired by previous axiomatic approaches \cite{Giles1964,
LiebYngvason1998} and discussions of Maxwell's demon \cite{Szilard1929,
Bennett1982}.  The basic concept of our system is the {\em eidostate}, which
is a collection of possible states from the point of view of some agent.
A review of our axioms and a few their consequences can be found in 
the Appendix.  The axioms imply the existence of additive conserved 
quantities called {\em components of content} and an entropy function
$\entropy$ that identifies reversible and irreversible processes.
The entropy includes both thermodynamic and information components.

One of the surprising things about this axiomatic system is that,
despite the absence of probabilistic ideas in the axioms, a concept
of probability emerges from the entropy $\entropy$.  If state $e$ is an
element of a uniform eidostate $E$, then we can define
\begin{equation}
	P(e|E) = \frac{2^{\entropy(e)}}{2^{\entropy(E)}} .
\end{equation}
States in $E$ with higher entropy are assigned higher probability.
As we will review below, this distribution has a uniquely simple
relationship to the entropies of the individual states and the overall
eidostate $E$.

The emergence of an entropic probability distribution motivates
us to ask several questions.  Can this idea be extended beyond
uniform eidostates?  Can we interpret an arbitrary probability
distribution over a set of states as an entropic distribution within
a wider context?  What does the entropic probability tell us about
probabilistic processes affecting the states within our axiomatic
system?  In this paper we will address these questions.

\section{Coin-and-box model}

We first review a few of the ideas of the system in \cite{AxiomaticInfoThermo}
by introducing a simple model of the axioms.  None of our later results
depend on this model, but a definite example will be convenient for 
explanatory purposes.  Our theory deals with configurations of coins
and boxes; as we will see below, the states are arrangements of coins,
memory records, and closed boxes containing coins.  
States are combined together using the $+$ operation, which simply
stands for ordered pairing of two states.  If $a \neq b$, $a+b$ is not the
same as $b+a$, and $a+(a+a)$ is distinct from $(a+a)+a$.  Thus,
the combination operation $+$ is neither commutative nor associative.

We construct our states from some elementary pieces:
\begin{itemize}
	\item  {\em Coin states}, which can be either $h$ (heads) or
		$t$ (tails) or combinations of these.  It is also convenient 
		to define a {\em stack state} $s_{n}$ to be a particular
		combination of $n$ coin states $h$:  
		$s_{n} = h + (h +(h + \cdots))$.
		The {\em coin value} $Q$ of a compound of coin states is
		just the total number of coins involved.  A finite set
		$K$ of coin states is said to be $Q$-uniform if every
		element has the same $Q$-value.
	\item  Record states $r$.  As the name suggests, 
		these should be interpreted as specific values
		in some available memory register.  The combination
		of two record states is another record state.  Thus,
		$r$, $r+r$, $r+(r+r)$, etc., are all distinct record states.
		Record states are not coins, so $Q(r) = 0$.
	\item  Box states.  For any $Q$-uniform set of coin
		states $C$, there is a sequence of {\em box states}
		$\boxstate{K}{n}$.  Intuitively, this represents a kind
		of closed box containing $n Q(K)$ coins, so that
		$Q\left ( \boxstate{K}{n} \right ) = n Q(K)$.  If 
		$K = \{ h,t \}$ then we denote the corresponding 
		``basic'' box states by $b_{n}$.
\end{itemize}
An {\em eidostate} is any finite, non-empty, $Q$-uniform set of states.
The $+$ operation on eidostates is just the Cartesian product
of the sets, and always yields another eidostate.  For convenience,
we identify the state $s$ with the singleton eidostate $\{ s \}$.

We now must define the relation $\rightarrow$, which tells
us which states can be transformed into which other states.
We will first give some elementary relations:
\begin{itemize}
	\item  Two eidostates are {\em similar} (written $A \sim B$) if
		they are composed of the same Cartesian factors, 
		perhaps combined in a different way.  If $A \sim B$, then
		$A \leftrightarrow B$.  (The notation $\leftrightarrow$ 
		means $A \rightarrow B$ and $B \rightarrow A$.)  As
		far as the $\rightarrow$ relation is concerned, we can
		freely rearrange the ``pieces'' in a compound eidostate.
	\item  For coin states, $h \leftrightarrow t$.
	\item  If $r$ is a record state, $a + r \leftrightarrow a$ for
		any $a$.  In a similar way, for an {\em empty} box
		state, $a + \boxstate{K}{0} \leftrightarrow a$.
	\item  If $K$ is a $Q$-uniform eidostate of coin states,
		$\boxstate{K}{n} + K \leftrightarrow \boxstate{K}{n+1}$.
\end{itemize}
Now we add some rules that allow us to extend these to 
more complex situations.  In what follows, $A$, $A'$, $B$, etc., 
are eidostates, and $s$ is a state.
\begin{description}
	\item[Transitivity.]  If $A \rightarrow B$ and $B \rightarrow C$, then $A \rightarrow C$.
	\item[Augmentation.]  If $A \rightarrow B$, then $A+C \rightarrow B+C$.
	\item[Cancelation.]  If $A+s \rightarrow B+s$, then $A \rightarrow B$.
	\item[Subset.]  If $A \rightarrow s$ and $A' \subseteq A$, then $A' \rightarrow s$.
	\item[Disjoint union.]  If $A$ and $B$ are both disjoint unions 
		$A = A_{1} \cup A_{2}$ and $B_{1} \cup B_{2}$, and both 
		$A_{1} \rightarrow B_{1}$ and $A_{2} \rightarrow B_{2}$, then
		$A \rightarrow B$.
\end{description}

Using these rules we can prove a lot of $\rightarrow$ relations.
For example, for a basic box state we have $b_{n} + \{ h,t \} \leftrightarrow b_{n+1}$.
From the subset rule we have $b_{n} + h \rightarrow b_{n+1}$
(but not the reverse).  Then we can say,
\begin{equation}  \label{eq-coinflip}
	b_{n} + h \rightarrow b_{n+1} \rightarrow b_{n} + \{ h,t \},
\end{equation}
from which we can conclude (via transitivity and cancelation)
that $h \rightarrow \{ h,t \}$.  The use of a basic box allows us
to ``randomize'' the state of one coin.

Or consider two coin states and distinct record states $r_{0}$ and $r_{1}$.
Then
\begin{equation}
	h \rightarrow h+r_{0} \qquad \mbox{and} \qquad t \rightarrow t+r_{1} \rightarrow h+r_{1},
\end{equation}
from which can show that $\{ h,t \} \rightarrow h + \{ r_{0}, r_{1} \}$.  That is,
we can set an unknown coin state to $h$, if we also make a record of 
which state it is.  A pretty similar argument establishes the following:
\begin{eqnarray}
	(h + \{ r_{0},r_{1} \}) + b_{n} & \rightarrow & \{ h+r_{0}, t+r_{1} \} + b_{n} \nonumber \\ 
	& \rightarrow & ( \{ h,t \} + r_{0}) + b_{n} \rightarrow \{ h,t \} + b_{n} \rightarrow b_{n+1} .
\end{eqnarray}
The eidostate $\{ r_{0}, r_{1} \}$ (called a {\em bit state}) can be deleted
at the cost of a coin absorbed by the basic box.  The basic box is a 
coin-operated deletion device; and since each step above is reversible,
we can also use it to dispense a coin together with a bit state 
(that is, an unknown bit in a memory register).

These examples help us to clarify an important distinction.  What is
the difference between the box state $b_{1}$ and the eidostate
$\{ h,t \}$?  Could we simply replace all box states $\boxstate{K}{n}$
with a simple combination $K + (K + \ldots)$ of possible coin eidostates?
We cannot, because such a replacement would preclude us from 
using the subset rule to obtain Equation~\ref{eq-coinflip}.  The whole
point of the box state is that the detailed state of its contents is
{\em entirely inaccessible} for determining possible processes.
Putting a coin in a box effectively randomizes it.  

It is not difficult to show that our model satisfies all 
of the axioms presented in the Appendix,
with the mechanical states in $\mechset$ identified as coin states.
The key idea in the proof is that we can reversibly reduce any 
eidostate to one with a special form:
\begin{equation}
	A \leftrightarrow s_{q} + I_{k} ,
\end{equation}
where $s_{q}$ is a stack state of $q = Q(A)$ coins and $I_{k}$
is an {\em information state} containing $k$ possible record states.
Relations between eidostates are thus reduced to relations
between states of this form.
We note that the coin value $q$ is conserved in every
$\rightarrow$ relation, and no relation allows us to decrease
the value of $k$.  In our model, there is just one independent 
component of content ($Q$ itself), 
and the entropy function is $\entropy(A) = \log k$.
(We use base-2 logarithms throughout.)

\section{The entropy formula and entropic probability}

Now let us return to the general axiomatic system.  A uniform
eidostate $E$ is one for which, given any two states $e,f \in E$,
either $e \rightarrow f$ or $f \rightarrow e$.  (We may write
this disjunction as $e \leftorright f$.)  The set of all
uniform eidostates is called $\uniformset$.   Then the axioms
imply the following theorem (Theorem 8 in \cite{AxiomaticInfoThermo}):
\begin{result}
	There exist an entropy function $\entropy$ and a set of 
	components of content $Q$ on $\uniformset$ with the following properties:
	\begin{description}
		\item(a)  For any $E,F \in \uniformset$, $\entropy(E+F) = \entropy(E) + \entropy(F)$.
		\item(b)  For any $E,F \in \uniformset$ and component of content $Q$, 
			$Q(E+F) = Q(E)+Q(F)$.
		\item(c)  For any $E,F \in \uniformset$, $E \rightarrow F$ if and only if
			$\entropy(E) \leq \entropy(F)$ and $Q(E) = Q(F)$ for every
			component of content~$Q$.
		\item(d)  $\entropy(m) = 0$ for all $m \in \mechset$.
	\end{description}
\end{result}
The entropy function $\entropy$ is determined\footnote{Up to a non-mechanical
component of content} by the $\rightarrow$ relations among the eidostates.

We can compute the entropy of a uniform eidostate $E$ in terms of the entropies
of its elements $e$.  This is
\begin{equation}
	\entropy(E) = \log \left (  \sum_{e \in E} 2^{\entropy(e)} \right ) .
\end{equation}
It is this equation that motivates our definition of the entropic probability of $e$
within the eidostate $E$:
\begin{equation}
		P(e|E) = \frac{2^{\entropy(e)}}{2^{\entropy(E)}} .
\end{equation}
Then $P(e|E) \geq 0$ and the probabilities sum over $E$ to 1.  
As we have mentioned, the entropy function $\entropy$ may not be
quite unique; nevertheless, two different admissible entropy functions
lead to the same entropic probability distribution.   Even better,
our definition gives us a very suggestive formula for the entropy of $E$:
\begin{eqnarray}
	\entropy(E) & = & \sum_{e \in E} P(e|E) \entropy(e) - \sum_{e \in E} P(e|E) \log P(e|E) \\
	& = & \bigg\langle \entropy(a) \bigg\rangle + H(\vec{P}),  \label{eq-avgplusshannon}
\end{eqnarray}
where the mean $\langle \cdots \rangle$ is taken with respect to the entropic
probability, and $H(\vec{P})$ is the Shannon entropy of the distribution $P$
\cite{Shannon1948,CoverThomas}.

Equation~\ref{eq-avgplusshannon} is very special.  If we choose an arbitrary
distribution (say $P'(e|E)$), then with respect to this probability we find
\begin{equation}
	\entropy(E) \geq \biggl \langle \entropy(a) \biggr \rangle_{\!\!P'} + H(\vec{P}'),
\end{equation}
with equality if and only if $P'$ is the entropic distribution \cite{CoverThomas}.  
Therefore we might {\em define} the entropic probability to be the distribution 
that maximizes the sum of average state entropy and Shannon entropy---a kind 
of ``maximum entropy'' characterization.

\section{Uniformization}

A unique entropic probability rule arises from our $\rightarrow$ relations among
eidostates, which in the real world might summarize empirical data about 
possible state transformations.  But so far, this entropic probability distribution
$P(e|E)$ is only defined within a uniform eidostate $E$.  

In part this makes sense.  An eidostate represents represents the knowledge of
an agent---i.e., that the state must be one of those included in the set.  
This is the knowledge upon which the agent will assign probabilities, which is
why we have indicated the eidostate $E$ as the {\em condition} for the distribution.
Furthermore, these might be the only eidostates, since the axioms themselves do not
guarantee that any non-uniform eidostates exist.
(Some models of the axioms have them, and some do not.)
But can we generalize the probabilities to distributions over non-uniform 
collections of states?

Suppose $A = \{ a, a', \ldots \}$ is a finite set of states, possibly not uniform.
Then we say that $A$ is {\em uniformizable} if there exists a uniform
eidostate $\hat{A} = \{ a + m_{a}, a'+m_{a'}, \ldots \}$, where the 
states $m_{a}$ are mechanical states in $\mechset$.  
The idea is that the states in $A$, which vary in their components of
content, can be extended by mechanical states that ``even out'' these
variations.  Since $\hat{A}$ is uniform, then 
$a + m_{a} \leftorright a'+m_{a'}$ for any
$a, a' \in A$.  The abstract process $\fproc{a}{a'}$ is said to be
{\em adiabatically possible} \cite{Giles1964}.
Mechanical states have $\entropy(m_{a}) = 0$, 
so the entropy of the extended
$\hat{A}$ is just
\begin{equation}
	\entropy(\hat{A}) = \log \left (  \sum_{a \in A} 2^{\entropy(a)} \right ),
\end{equation} 
which is independent of our choice of the uniformizing mechanical states.

What is the significance of this entropy?  Suppose $A$ and $B$ are not
themselves uniform, but their union $A \cup B$ is uniformizable.  Then
we may construct uniform eidostates $\hat{A} = \{ a+m_{a}, \ldots \}$ and 
$\hat{B} = \{ b+m_{b}, \ldots \}$ such that either $\hat{A} \rightarrow \hat{B}$
or $\hat{B} \rightarrow \hat{A}$, depending on whether 
$\entropy(\hat{A}) \geq \entropy(\hat{B})$ or the reverse.
In short, the entropies of the extended eidostates determine whether
the set of states $A$ can be turned into the set $B$, {\em if} we
imagine that these states can be augmented by mechanical states,
embedding them in a larger, uniform context.

Given the entropy of the extended state, we can define
\begin{equation}
	P(a|A) = P(a+m_{a}|\hat{A}) = \frac{2^{\entropy(a)}}{2^{\entropy(\hat{A})}} .
\end{equation}
This extends the entropic probability to the uniformizable set $A$.

Let us consider an example from our coin-and-box model.  We start out
with the non-uniform set $B = \{ b_{n}, b_{n+1} \}$.  These two basic
box states have different numbers of coins.  But we can uniformize
this set by adding stack states, so that $\hat{B} = \{ b_{n} + s_{m+1}, 
b_{n+1}+s_{m} \}$ is a uniform eidostate.  The entropy of a basic
box state is $\entropy(b_{n}) = n$,  so we have
\begin{equation}
	\entropy(\hat{B}) = \log \left ( 2^{n} + 2^{n+1} \right ) 
		= \log \left ( 3 \cdot 2^{n} \right )  = n + \log 3 .
\end{equation}
The entropic probabilities are thus
\begin{equation}  \label{eq-onethirdtwothirds}
	P(b_{n}|B) = \frac{1}{3}  \qquad \mbox{and} \qquad P(b_{n+1}|B) = \frac{2}{3} .
\end{equation}

\section{Reservoir states}

So far, we have uniformized a non-uniform set $A$ by augmenting
its elements with mechanical states, which act as a sort of ``reservoir''
of components of content.  These mechanical states have no entropy
of their own.  But we can also consider a procedure in which the 
augmenting states act more like the states of a thermal reservoir
in conventional thermodynamics.

We begin with a mechanical state $\mu$, and posit a sequence 
of {\em reservoir states} $\theta_{n}$, which have the following 
properties.
\begin{itemize}
	\item  For any $n$, $\theta_{n} + \mu \rightarrow \theta_{n+1}$.
	\item  $\theta_{k} + \theta_{l} \leftrightarrow \theta_{m} + \theta_{n}$
		if and only if $k+l = m+n$.
\end{itemize}
The reservoir states $\theta_{n}$ form a ladder.  We can ascend
one rung in the ladder by ``dissolving'' the mechanical state $\mu$
into the reservoir.  If we have more than one reservoir state, we can
ascend one ladder provided we descend another by the same number
of rungs.

For any $n$ and $m$, we have that $\entropy(\theta_{n}) + \entropy(\theta_{m+1}) 
= \entropy(\theta_{n+1}) + \entropy(\theta_{m})$,
so that
\begin{equation}
	\sigma = \entropy(\theta_{n+1}) - \entropy(\theta_{n}) = 
		\entropy(\theta_{m+1}) - \entropy(\theta_{m}) 
\end{equation}
is a non-negative constant for the particular sequence of reservoir 
states.  This sequence $\{ \theta_{n} \}$ is characterized by the state $\mu$ 
and the entropy increment $\sigma$.
Note that we can write $\entropy(\theta_{n}) = n \sigma + S_{0}$,
where $S_{0} = \entropy(\theta_{0})$. 

For example, in our coin-and-box model,
the basic box states $b_{n}$ act as a sequence of reservoir states
with a mechanical (coin) state $\mu = h$ and an entropy 
increment $\sigma = \log 2 = 1$.  The more general box
states $\boxstate{K}{n}$ form a reservoir state sequence
with $\mu = s_{q}$ and $\sigma = \log k$,
where $q = Q(K)$ and $k$ is the number of states in $K$.
For each of these bojx-state reservoir sequences, 
$S_{0} = \entropy(\theta_{0}) = 0$.

One particular type of reservoir is a {\em mechanical} reservoir
consisting of the states $\mu$, $\mu+\mu$, $\mu + (\mu + \mu)$, etc.
We denote the $n$th such state by $\mu_{n}$.  For the $\mu_{n}$ 
reservoir states, $\sigma = 0$.  If we have a finite set of states
$A = \{ a, a', \ldots \}$ that can be uniformized by the addition 
of the $\mu_{n}$ states, they can also be uniformized by a corresponding
set of non-mechanical reservoir states $\theta_{n}$:
\begin{equation}
	\hat{A} = \{ a + \theta_{n_{a}}, a' + \theta_{n_{a'}}, \ldots \} .
\end{equation}
As before, we can find the entropy of this uniform eidostate
and define entropic probabilities.  But the $\theta$ reservoir
states now contribute to the entropy and affect the 
probabilities.

First, the entropy:
\begin{equation}
	\entropy(\hat{A}) = \log \left ( \sum_{a \in A} 2^{\entropy(a) + \entropy(\theta_{n_{a}})} \right )
		= S_{0} + \log \left ( \sum_{a \in A} 2^{\entropy(a)} 2^{n_{a} \sigma} \right ) .
\end{equation}
The entropic probability---which now depends on the choice of reservoir 
states---is
\begin{equation}
	P_{\theta}(a|A) = P(a + \theta_{n_{a}} | \hat{A} )
	= \frac{2^{\entropy(a)+n_{a}\sigma}}{\displaystyle \left ( \sum_{a \in A} 2^{\entropy(a)+n_{a}\sigma} \right )} .
\end{equation}
The reservoir states affect the relative probabilities of the states.  For
example, suppose $\entropy(a) = \entropy(a')$ for a pair of states in $A$.
We might naively think that these states would end up with the same
entropic probability, as they would if we uniformized $A$ by mechanical
states.  But since we are uniformizing using the $\theta$ reservoir
states, it may be that $\theta_{n_{a}}$ and $\theta_{n_{a'}}$ have different
entropies.  Then the ratio of the probabilities is
\begin{equation}
	\frac{P_{\theta}(a|A)}{P_{\theta}(a'|A)} = \frac{2^{n_{a} \sigma}}{2^{n_{a'} \sigma}}
	= 2^{(n_{a}-n_{a'}) \sigma} ,
\end{equation}
which may be very different from 1.

Again, let us consider our coin-and-box model.  We begin with the non-uniform
set $A = \{ h, t, h+t \}$.  Each of these states has the same entropy $\entropy$,
that is, zero.  We choose to uniformize using basic box states $b_{n}$.
For instance, we might have
\begin{equation}
	\hat{A} = \{ h+b_{1}, t + b_{1}, (h+t) + b_{0} \}.
\end{equation}
Recalling that $\sigma = 1$, the entropy is
\begin{equation}
	\entropy(\hat{A}) = \log \left ( 2^{1} + 2^{1} + 2^{0} \right ) = \log 5 . 
\end{equation}
This yields probabilities
\begin{equation}
	P_{b}(h|A) =  \frac{2}{5} \qquad
	P_{b}(t|A)  = \frac{2}{5}  \qquad
	P_{b}(h+t|A) =  \frac{1}{5} .
\end{equation}

\begin{figure}
\begin{center}
	\includegraphics[width=1.75in]{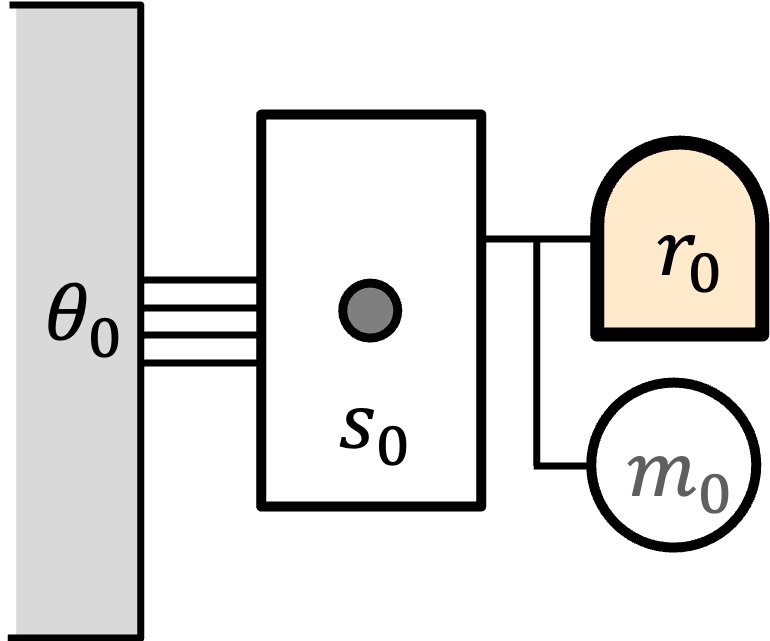}
	\caption{A simple Maxwell's demon.  \label{fig-demon1}}
\end{center}
\end{figure}

As an illustration of these ideas, consider the version of Maxwell's demon
shown in Figure~\ref{fig-demon1}.
The demon is a reversible computer with an initial memory state
$r_{0}$.  It is equipped with a reversible battery for storing energy,
initially in mechanical state $m_{0}$.  The demon interacts with 
a one-particle ``Szilard'' gas, 
in which the single particle can move freely within its volume 
(state $s_{0}$).  The gas is maintained in thermal equilibrium with
a heat reservoir, whose initial state is $\theta_{0}$.  We might
denote the overall initial state by $((r_{0} + m_{0})+s_{0}) + \theta_{0}$.

\begin{figure}
\begin{center}
	\includegraphics[width=5in]{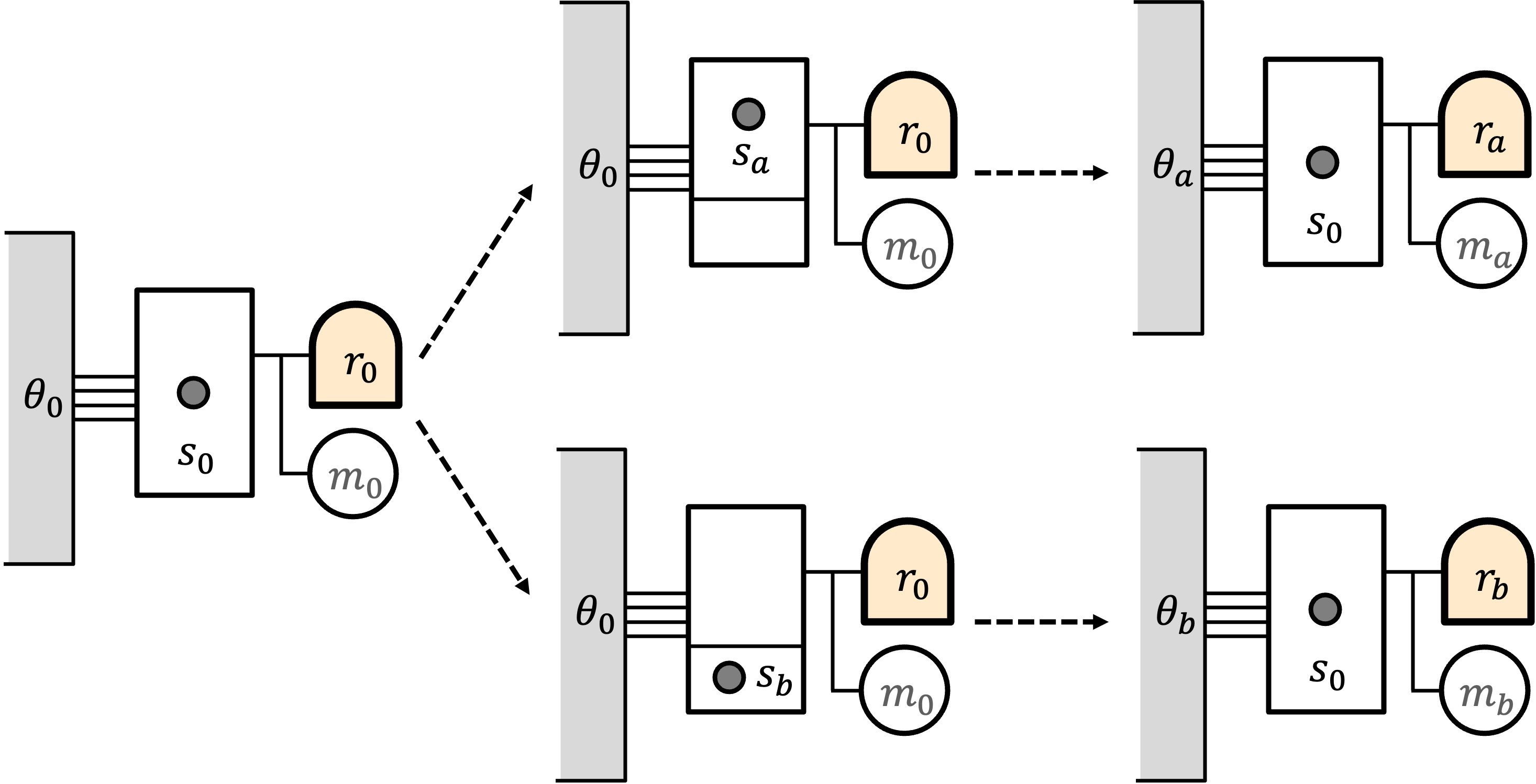}
	\caption{Extraction of work by dividing gas enclosure
	into unequal volumes..  \label{fig-demon2}}
\end{center}
\end{figure}

Now the demon introducers a partition into the gas, separating
the enclosure into unequal subvolumes, as in Figure~\ref{fig-demon2}.
The two resulting 
states are $s_{a}$ and $s_{b}$, which are not equally probable.
The probabilities here are entropic probabilities due to the 
difference in entropy of $s_{a}$ and $s_{b}$.
Now the demon records the location of the particle in its memory
and uses this to control the isothermal expansion of the one-particle
gas.  The work is stored in the battery.  
At the end of this process, the demon retains its memory record,
the battery is in one of two mechanical states $m_{a}$ and $m_{b}$.
The gas is again in state $s_{0}$.  But different amounts of heat have been
extracted from the reservoir during the expansion, so the reservoir
has two different states $\theta_{a}$ and $\theta_{b}$.

The overall final eidostate might be represented as
\begin{equation}
	F = \{ ((r_{a} + m_{a})+s_{0}) + \theta_{a},((r_{b} + m_{b})+s_{0}) + \theta_{b} \} .
\end{equation}
The states of the demon and the gas, $(r_{a} + m_{a})+s_{0}$ and 
$(r_{b} + m_{b})+s_{0}$, have different energies and the same entropy.
It is the reservoir states $\theta_{a}$ and $\theta_{b}$ that (1) make
$F$ uniform (constant energy), and (2) introduce the entropy differences
leading to different entropic probabilities for the two states.

A conventional view would suppose that the unequal probabilities for the
two final demon states comes from their history---that is, that the 
probabilities are inherited from the unequal partition of the gas.
In the entropic view, the unequal probabilities are due to differences
in the {\em environment} of the demon, represented by the different
reservoir states $\theta_{a}$ and $\theta_{b}$.  The environment, 
in effect, serves as the ``memory'' of the history of the process.

\section{Context states}

When we uniformize a non-uniform $A$ by means of a sequence of 
reservoir states, the reservoir states affect the entropic probabilities.
We can use this idea more generally.

For example, in our coin-and-box model, suppose we flip a coin but do not 
know how it lands.  This might be represented by the eidostate $F = \{ h, t \}$.
Without further information, we would assign the coin states equal probability
1/2, which is the simple entropic probability.  But suppose we have additional
information about the situation that would lead us to assign probabilities
1/3 and 2/3 to the coin states.  This additional information---this {\em context}---must
be reflected in the eidostate.  The example in Equation~\ref{eq-onethirdtwothirds}
tells us that this does the job:
\begin{equation}
	\hat{F} = \{ h + (b_{n} + s_{m+1}), t + (b_{n+1} + s_{m}) \} .
\end{equation}
The extended coin-flip state $\hat{F}$ includes extra context so 
that the entropic probability reflects our additional information.

In general, we can adjust our entropic probabilities by incorporating
{\em context states}.  Suppose we have a uniform eidostate 
$E = \{ e_{1}, e_{2}, \ldots \}$, but we wish to {\em specify} a 
particular non-entropic distribution $p_{k}$ over these states.  
Then for each $e_{k}$ we introduce eidostates $C_{k}$, 
leading to an extended eidostate
\begin{equation}
	\hat{E} = \bigcup_{k} \left ( e_{k} + C_{k} \right ) ,
\end{equation}
which we assume is uniform.  The $C_{k}$'s are the context 
states.  Our challenge is to find a set of context states 
so that the entropic probability in $\hat{E}$
equals the desired distribution $p_{k}$.

We cannot always do this exactly, but we can always approximate 
it as closely as we like.
First, we note that we can always choose our context eidostates
to be information states.  The information state $I_{n}$ containing
$n$ record states has entropy $\log n$.  Now for each $k$,
we closely approximate the ratio $p_{k} / 2^{\entropy(e_{k})}$ 
by a rational number; and since there are finitely many of these
numbers, we can represent them using a common denominator.  
In our approximation,
\begin{equation}
	\frac{p_{k}}{2^{\entropy(e_{k})}} = \frac{n_{k}}{N} .
\end{equation}
Now choose $C_{k} = I_{n_{k}}$ for each $k$.  The 
entropy of $\hat{E}$ becomes
\begin{eqnarray}
	\entropy(\hat{E}) & = & \log \left ( \sum_{k} 2^{\entropy(e_{k} + \log n_{k})} \right ) \nonumber \\
		& = & \log \left ( \sum_{k} n_{k} 2^{\entropy(e_{k})} \right )
		= \log \left ( \sum_{k} p_{k} N \right ) = \log N .
\end{eqnarray}
From this, we find that the entropic probability is
\begin{equation}
	P(e_{k}+I_{k} | \hat{E}) = \frac{n_{k} 2^{\entropy(e_{k})}}{N} = p_{k},
\end{equation}
as desired.

We find, therefore, that the introduction of context states $C_{k}$ 
allows us to ``tune'' the entropic probability to approximate any
distribution $p_{k}$ that we like.  This is more than a trick.  The
distribution $p_{k}$ represents additional implicit information 
(beyond the mere list of states $E = \{ e_{k} \}$), and such 
additional information must have a physical representation.  
The context states are that representation.

\section{Free energy}

The tools we have developed can lead to some interesting places.
Suppose we have two sets of states, $A=\{ a_{i} \}$ and 
$B=\{ b_{j} \}$, endowed with {\em a priori} probability 
distributions $p_{i}$ and $q_{j}$, respectively.  We wish to
know when the states in $A$ can be turned into the states
in $B$, perhaps augmented by reservoir states.  That is, we
wish to know when $\hat{A} \rightarrow \hat{B}$.

We suppose we have a mechanical state $\mu$, leading to a 
ladder of mechanical reservoir states $\mu_{n} = \mu + (\mu + \ldots)$.
The mechanical state $\mu$ is non-trivial, in the the sense that
$s + \mu \not \rightarrow s$ for any $s$.  This means that there is
a component of content $Q$ such that $Q(\mu) \neq 0$.  The set
$A \cup B$ can be uniformized by augmenting the $a_{i}$ and 
$b_{j}$ states by $\mu_{n}$ mechanical reservoir states.

However, we still need to realize the $p_{i}$ and $q_{j}$ 
probabilities.  We do this by introducing as context states a 
corresponding ladder of reservoir states $\theta_{n}$ such that
$\sigma = \entropy(\theta_{n+1}) - \entropy(\theta_{n})$ is
very small.  
Essentially, we assume that the reservoir states are 
``fine-grained'' enough that we can approximate any 
positive number by $2^{n \sigma}$ for some positive or
negative integer $n$.  Then, if we augment the $a_{i}$ and $b_{j}$ 
states by combinations of $\mu_{n}$ and $\theta_{n}$ states,
we can uniformize $A \cup B$ and also tune the entropic
probabilities to match the {\em a priori} $p_{i}$ and $q_{j}$.
The final overall uniform eidostate is
\begin{equation}
	\{ a_{i} + (\mu_{l_{i}} + \theta_{k_{i}}) , b_{j} + (\mu_{h_{j}} + \theta_{n_{j}}) \} ,
\end{equation}
for integers $l_{i}$, $k_{i}$, $h_{j}$ and $n_{j}$.  The uniformized
$\hat{A}$ and $\hat{B}$ eidostates are subsets of this, and thus
are themselves uniform eidostates.  The entropic probabilities have
been adjusted so that
\begin{equation}
	p_{i} = \frac{2^{\entropy(a_{i}) + k_{i} \sigma + \entropy(\theta_{0})}}{2^{\entropy(\hat{A})}}
	\quad \mbox{and} \quad
	q_{j} = \frac{2^{\entropy(b_{j}) + n_{j} \sigma + \entropy(\theta_{0})}}{2^{\entropy(\hat{B})}} .
\end{equation}
We now choose a component of content $Q$ such that $Q(\mu) = \varepsilon > 0$.
Since the overall state is is uniform, it must be true that
\begin{equation}
	Q(a_{i}) + l_{i} \varepsilon + k_{i} \varepsilon = Q(b_{j}) + h_{j} \varepsilon + n_{j} \varepsilon = \mbox{constant}
\end{equation}
for all choices of $i,j$.  Of course, if all of these values are the same, 
we can average them together and obtain
\begin{equation}
	\bigl \langle Q(a_{i}) \bigr \rangle_{\!p} + \bigl \langle l_{i} \bigr \rangle_{\!p} \varepsilon
	+ \bigl \langle k_{i} \bigr \rangle_{\!p} \varepsilon
	=
	\bigl \langle Q(b_{j}) \bigr \rangle_{\!q} + \bigl \langle h_{j} \bigr \rangle_{\!q} \varepsilon
	+ \bigl \langle n_{j} \bigr \rangle_{\!q} \varepsilon .
\end{equation}
We can write the average change in the $Q$-value of the mechanical state as
\begin{equation}  \label{eq-changeinmech}
	\left (  \bigl \langle h_{j} \bigr \rangle_{\!q} - \bigl \langle l_{i} \bigr \rangle_{\!p} \right ) \varepsilon
	= \left (  \bigl \langle k_{i} \bigr \rangle_{\!p} -  \bigl \langle n_{j} \bigr \rangle_{\!q} \right ) \varepsilon + 
	\bigl \langle Q(a_{i}) \bigr \rangle_{\!p} - \bigl \langle Q(b_{j}) \bigr \rangle_{\!q} .
\end{equation}

Since all of the states lie within the same uniform eidostate, 
$\hat{A} \rightarrow \hat{B}$ if and only if $\entropy(\hat{A}) \leq \entropy(\hat{B})$---that is,
\begin{equation}
	H(\vec{p}) + \bigl \langle \entropy(a_{i}) \bigr \rangle_{\!p} + \bigl \langle k_{i} \bigr \rangle_{\!p} \sigma
	\leq
	H(\vec{q}) + \bigl \langle \entropy(b_{j}) \bigr \rangle_{\!q} + \bigl \langle n_{j} \bigr \rangle_{\!q} \sigma .
\end{equation}
From this it follows that
\begin{equation}
	\left ( \bigl \langle k_{i} \bigr \rangle_{\!p} - \bigl \langle n_{j} \bigr \rangle_{\!q} \right ) \leq
	\frac {1}{\sigma} \left (   H(\vec{q}) - H(\vec{p}) + \bigl \langle \entropy(b_{j}) \bigr \rangle_{\!q}
	- \bigl \langle \entropy(a_{i}) \bigr \rangle_{\!p} \right ) .
\end{equation}
If we substitute this inequality into Equation~\ref{eq-changeinmech}, we obtain
\begin{eqnarray}  \label{eq-freeinequality1}
	\left (  \bigl \langle h_{j} \bigr \rangle_{\!q} - \bigl \langle l_{i} \bigr \rangle_{\!p} \right ) \varepsilon -
	\frac{\varepsilon}{\sigma} \left ( H(\vec{q}) - H(\vec{p}) \right ) & \leq &
	 + \frac{\varepsilon}{\sigma} \left ( \bigl \langle \entropy(b_{j}) \bigr \rangle_{\!q} - 
			\bigl \langle \entropy(a_{i}) \bigr \rangle_{\!p} \right ) \\ & & 
	- \left ( \bigl \langle Q(b_{j}) \bigr \rangle_{\!q} - \bigl \langle Q(a_{i}) \bigr \rangle_{\!p} \right ) . \nonumber
\end{eqnarray}

We can get insight into this expression as follows.  Given the process $\hat{A} \rightarrow \hat{B}$,
\begin{itemize}
	\item  $\left (  \bigl \langle h_{j} \bigr \rangle_{\!q} - \bigl \langle l_{i} \bigr \rangle_{\!p} \right ) \varepsilon$ is
		the average increase in $Q$-value of the mechanical state, which we can call $\avg{\Delta Q_{\mu}}$.
		Intuitively, this might be regarded as the ``work'' stored in the $\hat{A} \rightarrow \hat{B}$ process.
	\item  We can denote the change in the Shannon entropy of the probabilities by 
		$\Delta H = H(\vec{q}) - H(\vec{p})$.  Since each $a_{i}$ or $b_{j}$ state could be
		augmented by a corresponding record state, this is the change in the information
		entropy of the stored record.
	\item  For each state $a$, we can define the {\em free energy} 
		$F(a) = Q(a) - \frac{\varepsilon}{\sigma} \entropy(a)$.  We call this free ``energy'', even
		though $Q$ does not necessarily represent energy, because of the analogy with the
		familiar expression $F = E-TS$ for the Helmholtz free energy in conventional thermodynamics.
		The average change in the free energy $F$ is 
		\begin{equation}
			\avg{\Delta F} = \left ( \bigl \langle Q(b_{j}) \bigr \rangle_{\!q} - \bigl \langle Q(a_{i}) \bigr \rangle_{\!p} \right )
				-  \frac{\varepsilon}{\sigma} \left ( \bigl \langle \entropy(b_{j}) \bigr \rangle_{\!q} - 
				\bigl \langle \entropy(a_{i}) \bigr \rangle_{\!p} \right ) .
		\end{equation}
		The free energy $F$ depends on the particular reservoir states
		$\theta_{n}$ only via the ratio $\varepsilon/\sigma$.  Given this value,
		$\avg{\Delta F}$ depends only on the $a_{i}$ and $b_{j}$ states,
		together with their {\em a priori} probabilities.
		
		To return to our coin-and-box example, suppose we use the 
		basic box states $b_{n}$ as reservoir states $\theta_{n}$,
		and we choose the coin number $Q$ as our component of content. 
		Then $\varepsilon = 1$ and $\sigma = 1$, so that the free energy
		function $F(a) = Q(a) - \entropy(a)$.  (If we use different
		box states $\boxstate{K}{n}$ as reservoir states, the ratio
		$\varepsilon/\sigma$ is different.)
\end{itemize}
With these definitions, Equation~\ref{eq-freeinequality1} becomes
\begin{equation}
	\avg{\Delta Q_{\mu}} - \frac{\varepsilon}{\sigma} \Delta H \leq - \avg{\Delta F} .
\end{equation}
Increases in the average stored mechanical work, and decreases in the stored information,
must be paid for by a corresponding decrease in the average free energy.

Many useful inferences can be drawn from this.  For example, the erasure $Q$-cost of
one bit of information in the presence of the $\theta$-reservoir is $\varepsilon/\sigma$.
This cost can be paid from either the mechanical $Q$-reservoir state or the average 
free energy, or from a combination of these.  This amounts to a very general
version of Landauer's principle \cite{Landauer1961}, one that involves any type of mechanical component 
of content.

\section*{Appendix}

In this appendix we review some of the main definitions and axioms of the
theory, as well as some of its key results.  For more details, please
see \cite{AxiomaticInfoThermo}.

The theory is built a few essential elements:
\begin{itemize}
	\item  A set $\stateset$ of states and a set $\eidoset$ of eidostates.
		Each eidostate is a finite collection of states.  Without too
		much confusion, we may identify a state $a \in \stateset$
		with the singleton eidostate $\{ a \} \in \eidoset$, so that
		$\stateset$ can be regarded as a subset of $\eidoset$.
	\item  An operation $+$ by which eidostates are combined.  This
		is just the Cartesian product of the sets.  Two eidostates
		$A$ and $B$ are similar ($A \sim B$) if they are formed by the
		same Cartesian factors, perhaps put together in a different way.
	\item  A relation $\rightarrow$ on $\eidoset$.  We interpret $A \rightarrow B$
		to mean that eidostate $A$ may be transformed into eidostate $B$.
		A process is a pair $\fproc{A}{B}$ of eidostates, and it is said to 
		be possible if either $A \rightarrow B$ or $B \rightarrow A$.  An
		eidostate $A$ is uniform if, for all $a,b \in A$, $\fproc{a}{b}$ is possible.
	\item  Special states in $\stateset$ called record states.  State $r$ is a 
		record state if there exists another state such that $a + r \leftrightarrow a$.
		An information state is an eidostate containing only record states; the
		set of these is called $\infoset$.  A bit state $\bitstate$ is an information 
		state with exactly two elements, and a bit process is a process of the
		form $\fproc{r}{I_{b}}$.
\end{itemize}

Given this background, we can present our axioms.

\begin{description}
\item[Axiom I] (Eidostates.)
	$\eidoset$ is a collection of sets called \emph{eidostates} such that:
	\begin{description}
		\item[(a)]  Every $A \in \eidoset$ is a finite nonempty set
			with a finite prime Cartesian factorization.
		\item[(b)]  $A + B \in \eidoset$ if and only if $A,B \in \eidoset$.
		\item[(c)]  Every nonempty subset of an eidostate is also an eidostate.
	\end{description}
\item[Axiom II] (Processes.)
Let eidostates $A,B,C \in \eidoset$, and $s \in \stateset$.
\begin{description}
	\item[(a)]  If $A \sim B$, then $A \rightarrow B$.
	\item[(b)]  If $A \rightarrow B$ and $B \rightarrow C$, then $A \rightarrow C$.
	\item[(c)]  If $A \rightarrow B$, then $A + C \rightarrow B + C$.
	\item[(d)]  If $A + s \rightarrow B + s$, then $A \rightarrow B$.
\end{description}
\item[Axiom III]
	If $A,B \in \eidoset$ and $B$ is a proper subset of $A$, then $A \nrightarrow B$.
\item[Axiom IV] (Conditional processes.)
	\begin{description}
		\item[(a)]  Suppose $A, A' \in \eidoset$ and $b \in \stateset$.  If $A \rightarrow b$
			and $A' \subseteq A$ then $A' \rightarrow b$.
		\item[(b)]  Suppose $A$ and $B$ are uniform eidostates that are each disjoint
			unions of eidostates: $A = A_{1} \cup A_{2}$ and $B = B_{1} \cup B_{2}$.
			If $A_{1} \rightarrow B_{1}$ and $A_{2} \rightarrow B_{2}$ then
			$A \rightarrow B$.
	\end{description}
\item[Axiom V]  (Information.)
	There exist a bit state and a possible bit process.
\item[Axiom VI]  (Demons.)
	Suppose $a,b \in \stateset$ and $J \in \infoset$ such that $a \rightarrow b + J$.
	\begin{description}
		\item[(a)]  There exists $I \in \infoset$ such that $b \rightarrow a+I$.
		\item[(b)]  For any $I \in \infoset$, either $a \rightarrow b + I$ or $b+I \rightarrow a$.
	\end{description} 
\item[Axiom VII]  (Stability.)
	Suppose $A,B \in \eidoset$ and $J \in \infoset$.  If $nA \rightarrow nB + J$ for
	arbitrarily large values of $n$, then $A \rightarrow B$.
\item[Axiom VIII]  (Mechanical states.)
	There exists a subset $\mechset \subseteq \stateset$ of 
	\emph{mechanical states} such that:
	\begin{description}
		\item[(a)]  If $l,m \in \mechset$, then $l+m \in \mechset$.
		\item[(b)]  For $l,m \in \mechset$, if $l \rightarrow m$
			then $m \rightarrow l$.
	\end{description}	
\item[Axiom IX]  (State equivalence.)
	If $E$ is a uniform eidostate then there exist states $e,x,y \in \stateset$
	such that $x \rightarrow y$ and $E + x \leftrightarrow e + y$.
\end{description}

A component of content $Q$ is a real-valued additive function
on the set of states $\stateset$.  (Additive in this context means 
that $Q(a+b) = Q(a) + Q(b)$.)  Components of content
represent quantities that are conserved in every possible process.
In a uniform eidostate $E$, every element has the same values of
all components of content, so we can without ambiguity refer to
the value $Q(E)$.  The set of uniform eidostates is denoted $\uniformset$.
This set includes all singleton states in $\stateset$, all information states
in $\infoset$ and so forth, and it is closed under the $+$ operation.

\bibliographystyle{unsrt}
\bibliography{entropy}

\begin{thebibliography}{1}

\bibitem{AxiomaticInfoThermo}
Austin Hulse, Benjamin Schumacher, and Michael~D. Westmoreland.
\newblock Axiomatic information thermodynamics.
\newblock {\em Entropy}, 20(4):237, 2018.

\bibitem{Giles1964}
R.~Giles.
\newblock {\em Mathematical Foundations of Thermodynamics}.
\newblock Pergamon Press Ltd., Oxford, 1964.

\bibitem{LiebYngvason1998}
Elliott~H. Lieb and Jakob Yngvason.
\newblock A guide to entropy and the second law of thermodynamics.
\newblock {\em Notices of the American Mathematical Society}, 45:571--581,
  1998.

\bibitem{Szilard1929}
Leo Szilard.
\newblock On the decrease of entropy in a thermodynamic system by the
  intervention of intelligent beings.
\newblock {\em Zeitschrift fur Physik}, 53:840--856, 1929.
\newblock (English translation in {\em Behavioral Science} {\bf 1964}, 9,
  301--310.).

\bibitem{Bennett1982}
Charles~H. Bennett.
\newblock The thermodynamics of computation---a review.
\newblock {\em International Journal of Theoretical Physics}, 21:905--940,
  1982.

\bibitem{Shannon1948}
C.~E. Shannon.
\newblock A mathematical theory of communication.
\newblock {\em Bell System Technical Journal}, 27:379--423,623--656, 1948.

\bibitem{CoverThomas}
Thomas~M. Cover and Joy~A. Thomas.
\newblock {\em Elements of Information Theory (Second Edition)}.
\newblock John Wiley and Sons, Hoboken, 2006.

\bibitem{Landauer1961}
R.~Landauer.
\newblock Irreversibility and heat generation in the computing process.
\newblock {\em IBM Journal of Research and Development}, 5:183--191, 1961.

\end{thebibliography}

\end{document}